\documentclass[preprint,12pt]{elsarticle}
\usepackage{graphicx}
\usepackage{subfigure}
\usepackage{amssymb}

\journal{Physica A}

\begin{document}

\begin{frontmatter}
 
\title{Absence of phase transition in the XY-model on Menger sponge}

\author{M.A. Przedborski}
\author{B. Mitrovi\' c\corref{cor1}}
\ead{mitrovic@brocku.ca}
\cortext[cor1]{Corresponding author}
\address{Department of Physics, Brock University, St. Catharines, ON L2S 3A1, Canada}


\begin{abstract}
We have performed a Monte Carlo study of the classical XY-model on a Menger sponge with the Wolff cluster 
algorithm (U. Wolff, 1989). The Menger sponge is a fractal object with infinite order of ramification and fractal 
dimension $D = \log(20)/\log(3) = 2.7268$. From the dependence of the helicity modulus on system size and on 
boundary conditions, we conclude that there is no phase transition in the system at any finite temperature. \newline
\end{abstract}

\end{frontmatter}

\newpage

\section{Introduction}
In our previous Monte Carlo (MC) studies of the classical XY-model on Sierpi\' nski  
gasket with $D=$ 1.585~\cite{mb10}, and on Sierpi\' nski pyramid with fractal dimension $D = 2$~\cite{pm11}, 
we observed the absence of any finite temperature phase transition. These fractal lattices both have a
finite minimum order of ramification, $R_{min}$. The Sierpi\' nski pyramid has $R_{min}=$ 4, while
Sierpi\' nski gasket has $R_{min}=$ 3. The order of ramification $R$ at a point P of a lattice 
is the number of significant bonds one must break to 
isolate an arbitrarily large cluster of points surrounding P from the rest of the lattice. 
Regular, translationally invariant systems, which fill space uniformly, have infinite order
of ramification. In contrast, fractals, which are by definition scale invariant but not 
translationally invariant, do not fill space uniformly and may have a finite order of ramification.
In fractals with finite $R$, any cluster of points can be cut off from the rest of the structure by 
breaking only a finite number of bonds. Consequently, thermal fluctuations in these systems are 
sufficient to destroy the long range order associated with continuous phase transitions, as well as 
the quasi-long range order associated with the Berezinskii-Kosterlitz-Thouless (BKT) transition,  
exhibited by the XY-model in translationally invariant 
two-dimensional systems. This behavior is independent of the microscopic model and has been observed for 
spin models with discrete symmetry (Ising model, discrete $Z_{2}$ symmetry)~\cite{gab80,gab83,gasb84,gab84} 
as well as models with continuous symmetry (XY-model, continuous $O$(2) symmetry)~\cite{mb10,pm11,vkb91}. 

Gefen {\it et al.}~\cite{gab84} briefly explored the critical behavior of magnetic models with continuous 
$O(n)$ symmetry, $n\geq$ 2, on a class of Sierpi\' nski carpets,
which are two-dimensional analogues of Menger sponge, and are infinitely ramified. 
They used a correspondence between electrical properties of a pure resistor network connecting 
the sites of a lattice and the low-temperature properties of such magnetic models on the same lattice. 
They established the absence of long range order at finite temperatures in the 
continuous spin models on fractal structures with $D<$ 2, even in the case of an infinite $R$. They speculated
that this absence of long range order could be attributed to the fact that the lower critical dimension for 
spin models with continuous $O(n)$ symmetry on regular lattices is $d=$ 2. They remarked that 
the $O(n)$-model should be examined on fractals with $R = \infty$ and $D>$ 2. Recently, we have examined the 
classical XY-model on a Sierpi\' nski carpet~\cite{mp14}, 
which is a fractal with $R = \infty$ and $D=$ 1.8928, and we 
found no finite-temperature BKT transition to quasi-long-range order, in agreement with the conjecture of 
Gefen {\it et al.} 
 
\begin{figure}
\begin{center}
  \includegraphics[angle=0,width=4cm]{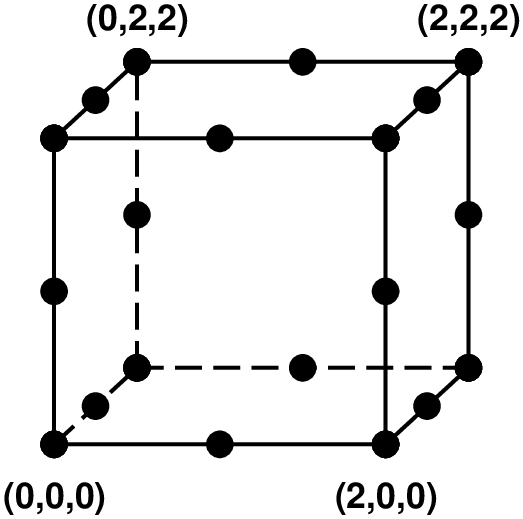}
\end{center}
\caption{The Menger sponge of order $m=$ 1.}
\label{fig:sponge1}
\end{figure}

Here we  examine the classical XY-model on a three-dimensional Menger sponge (MS), 
Figure~\ref{fig:sponge1}, which is a fractal structure that has both an infinite $R$ and fractal dimension  
$D>$ 2. The fractal dimension of MS is obtained from the scaling procedure which is used to construct it. 
One direct iterative process to generate MS begins with the first order ($m=$ 1) MS shown in 
Figure~\ref{fig:sponge1}, 
which is obtained by removing from the 2$\times$2$\times$2 structure the central sites of the six faces and the central 
site with coordinates (1,1,1). The second order ($m=$ 2) MS is then created by translating the first  
order sponge with 20 translation vectors, which are obtained by multiplying the position vectors of the 20 sites 
of the first order sponge by a factor of 3.  
The third order ($m=$ 3) MS is obtained by translating the second order sponge with 20 translation 
vectors which are obtained by increasing the length of the previous set of translation vectors by a factor of 3, etc. 
Each iteration in this process produces a structure with 20 times more sites than the previous one and with the edge 
which is three times longer than the edge of the previous structure. 
From the definition of the fractal dimension $D=\log($number of self-similar pieces$)/
\log($magnification factor$)$ one finds $D=\log($20$)/\log($3$)=$ 2.7268 for the MS. 

Similar to a cubic lattice ($R = \infty$), the number of bonds that must be cut to isolate an arbitrarily large 
cluster of points from the rest of the MS grows as a power of the linear size of the system. 
Consequently, the MS has an infinite order of ramification.  

Since MS is a fractal object, it is
scale invariant but not translationally invariant. The deviation from translational symmetry, as well as
the degree of inhomogeneity in the lattice, is measured by the lacunarity $L$. We used a procedure introduced by Gefen 
{\it et al.}~\cite{gab84} to estimate the lacunarity of our fractal, and found it to be $L=$ 0.19204~\cite{lac}. 

The Hamiltonian describing our system is given by
\begin{equation}
 H = -J\sum_{\langle i,j\rangle}\cos(\theta_{i}-\theta_{j})\>,
\label{eqn1}
\end{equation}
\noindent where $J>$ 0 is the coupling constant, $\langle i,j\rangle$ denotes the nearest neighbors, and 
0 $\leq \theta_{i}<$ 2$\pi$ is the angle variable on site $i$. We applied the Wolff cluster algorithm~\cite{wolff} to
perform a Monte Carlo study of Eq.~(\ref{eqn1}) on MS of several sizes. In our simulations, we computed  
the heat capacity, magnetic susceptibility, and helicity modulus for MS of order $m=$ 1--4 
(the number of sites in a sponge of order $m$ is $20^{m}$). The dependence of the helicity modulus on 
the boundary conditions and on system size indicates no finite temperature transition, despite the fact that MS is 
infinitely ramified and its fractal dimension exceeds the lower critical dimension of the XY-model.  

The remainder of
the paper is organized as follows. In Section~\ref{procedure} we present our numerical procedure used in   
calculations. In Section~\ref{results} we discuss our results, and summarize our findings in Section~\ref{summary}.  

\section{Calculation}\label{procedure}

One procedure for obtaining the MS of order $m$ was described in the Introduction: the sponge of order $m$ is
generated by translating the sponge of order $(m-$ 1$)$ with 20 translation vectors which are obtained by 
multiplying the position vectors of the 20 points in the $m=$ 1 sponge, shown in Figure 1, by a factor of 3$^{m-1}$.    
The XY-model described by Eq.~(\ref{eqn1}) takes into account only the nearest neighbor interactions, and it is  
necessary to provide a list of nearest neighbors for each spin on the lattice. By construction of the MS, 
two sites are nearest neighbors if their distance is equal to 1.  
It is important to note that since the MS is a fractal object and does not fill space uniformly,  
the average coordination number varies from one order to another. The sponges of order $m=$ 1, 2, 3, 4, and 5 
have average coordination numbers 2.400, 3.360, 3.744, 3.898, and 3.959, respectively, with standard deviations  
0.4899, 0.9113, 1.0181, 1.0532, and 1.0659, respectively. 
Because of the cubic symmetry of the system, the upper bound on the number of nearest neighbors for any 
point in the MS is equal to the coordination number of a simple cubic lattice. 

Since the fractals lack translational symmetry, we could not employ periodic boundary conditions.  
Instead we used only the closed boundary condition (CBC) and open (or free) boundary condition (OBC). 
For the closed boundary condition, each of the eight outer corners of the sponge of order $m$ is coupled  
to the three closest outer corners, which are located a distance of 3$^{m}$-1 away. For the open
boundary condition, the eight outer corners of the sponge are not coupled to each other.

In our simulations we 
used the Wolff cluster Monte Carlo algorithm~\cite{wolff} because it reduces the problems 
associated with the critical slowing down observed in the vicinity of a phase transition due to its 
non-local update scheme. In the Wolff cluster algorithm one starts building a cluster from a randomly 
chosen site $i$ in the lattice/sponge. At the same time one picks a randomly chosen axis through the origin 
of the two-dimensional classical planar spin space. All neighboring sites $j$ of $i$ are visited and these 
sites join the cluster with the probability
\begin{equation}
    \label{eq:prob}
    p_{ij} = 1 - \exp\left[-\frac{J}{k_B T}(s_i^{\perp} s_j^{\perp} + |s_i^{\perp} s_j^{\perp}|)\right],
\end{equation} 
where $s_i^{\perp}$ is the component of the spin on site $i$ which is perpendicular to the chosen axis, 
$k_B$ is the Boltzmann constant, and $T$ is the absolute temperature. Subsequently, one visits the neighbors $k$ 
of the new sites in the cluster, adding them to the cluster with probability $p_{jk}$ unless they are already in 
the cluster. This process is repeated until no new sites enter the cluster, at which point one flips the sign of 
all $s^{\perp}$ in the cluster, i.e. all spins in the cluster are  mirror-reflected in the chosen axis. 
This completes one Monte Carlo (MC) step in the Wolff algorithm and is analogous to a random change of angle of a 
single planar spin in the Metropolis MC algorithm~\cite{metropolis}. It is clear from Eq.~(\ref{eq:prob}) that 
the average size of clusters of correlated spins increases with decreasing temperature and could approach 
the system size at sufficiently low temperatures. Hence, the computing time at low temperatures rapidly increases  
with increasing order of the MS, and we considered only the sponges of order $m\leq$ 4.

For sponges of order $m\leq$ 3 
the simulations typically started at low temperatures and the initial configuration was a ``cold'' start, 
such that the system was in its lowest energy state with all the spins aligned. The final configuration at a given 
temperature was used as the initial configuration for the next higher temperature.  
For each temperature, we discarded the first 120,000 MC steps (clusters) to allow the system to equilibrate. An  
additional 7 links, each of 120,000 MC steps, was then generated at each temperature. To get an estimate of the error, 
we broke each of the 7 links into blocks of 20,000 MC steps. Then the average value was calculated for each of 
these 42 blocks, and the standard deviation of these 42 averages was used as an estimate of the error. 

The largest sponge ($m=$ 4 with 160,000 sites) was an exception to this procedure.  
Seven links of 120,000 MC steps per temperature required more than a month of computing time per temperature  
for $k_{B}T/J<$ 1.06. For the largest sponge we
started at the highest temperature $k_{B}T/J=$ 1.4 using a ``hot'' start, with the spins randomized and the 
final configuration was used as the initial configuration for the next lower temperature. As in the case of 
smaller sponges ($m<$ 4), the first 120,000 MC steps (clusters) were discarded and additional 7 links,  
each of 120,000 MC steps, were generated at each temperature. The lowest temperature at which the data were 
obtained in this way was $k_{B}T/J=$ 1.06, and the corresponding CPU time was one month. We obtained an 
additional set of data at $k_{B}T/J=$ 0.8 by reducing the size of link from 120,000 MC steps to 
30,000 MC steps per link and it took three weeks of CPU time to complete the simulation. 

The heat capacity per site was calculated using the fluctuations in internal energy
\begin{equation} 
  \label{eq:Cv}
 C = \frac{1}{N}\frac{\left<H^{2}\right>-\left<H\right>^{2}}{k_{B}T^{2}},
\end{equation}
and the linear magnetic susceptibility per site $\chi$ was calculated from fluctuations in 
magnetization per site ($m$) as
\begin{equation}
  \label{eq:chi}
 \chi=\frac{\left< m^{2} \right>- \left< m \right>^{2}}{k_{B}T},
\end{equation}
where $\left < \cdots \right >$ denotes the MC average.

The helicity modulus for the MS was calculated using the method of Shih, Ebner and Stroud~\cite{ses84} which 
we used in our previous work~\cite{mb10,pm11,mp14}. In this scheme, the XY Hamiltonian (Eq.~(\ref{eqn1})) is 
thought of as describing a set of Josephson-coupled superconducting grains in zero magnetic field. $\theta_{i}$, 
the direction of spin for each lattice site $i$, becomes the phase of the superconducting order parameter for 
that particular site. Applying a uniform vector potential $\bf{A}$ causes a shift in the phase difference 
$\theta_{i}-\theta_{j}$ of the XY Hamiltonian by the amount $2\pi{\bf A}\cdot({\bf r}_{j}-{\bf r}_{i})/\Phi_{0}$. 
In this expression, ${\bf r}_{i}$ is the position vector of site $i$, and $\Phi_{0}=hc/2e$ is the flux quantum. 
The helicity  
modulus is then obtained from the second derivative of the Helmholtz free energy per site with
respect to uniform $\bf{A}$, at $\bf{A}=0$. The resulting expression for $\gamma$ is
\begin{equation}
 \gamma=\frac{1}{N}\left[\left<\left(\frac{\partial^{2}H}{\partial A^{2}}\right)_{A=0}\right> 
-\frac{1}{k_{B}T}\left<\left(\frac{\partial H}{\partial A}\right)^{2}_{A=0}\right>
+\frac{1}{k_{B}T}\left<\left(\frac{\partial H}{\partial A}\right)_{A=0}\right>^{2}\right]\>,
\end{equation}
with H the phase-shifted XY Hamiltonian. The vector potential was applied along one of the cube edges of 
the MS (e.g. x-axis), which amounted to contributions to $\gamma$ only from nearest neighbors whose 
x-coordinates differ by $\pm1$ because of the cubic symmetry of the MS.

\section{Results and Discussion}\label{results}

\begin{figure}
\begin{center}
  \includegraphics[angle=0,width=8cm]{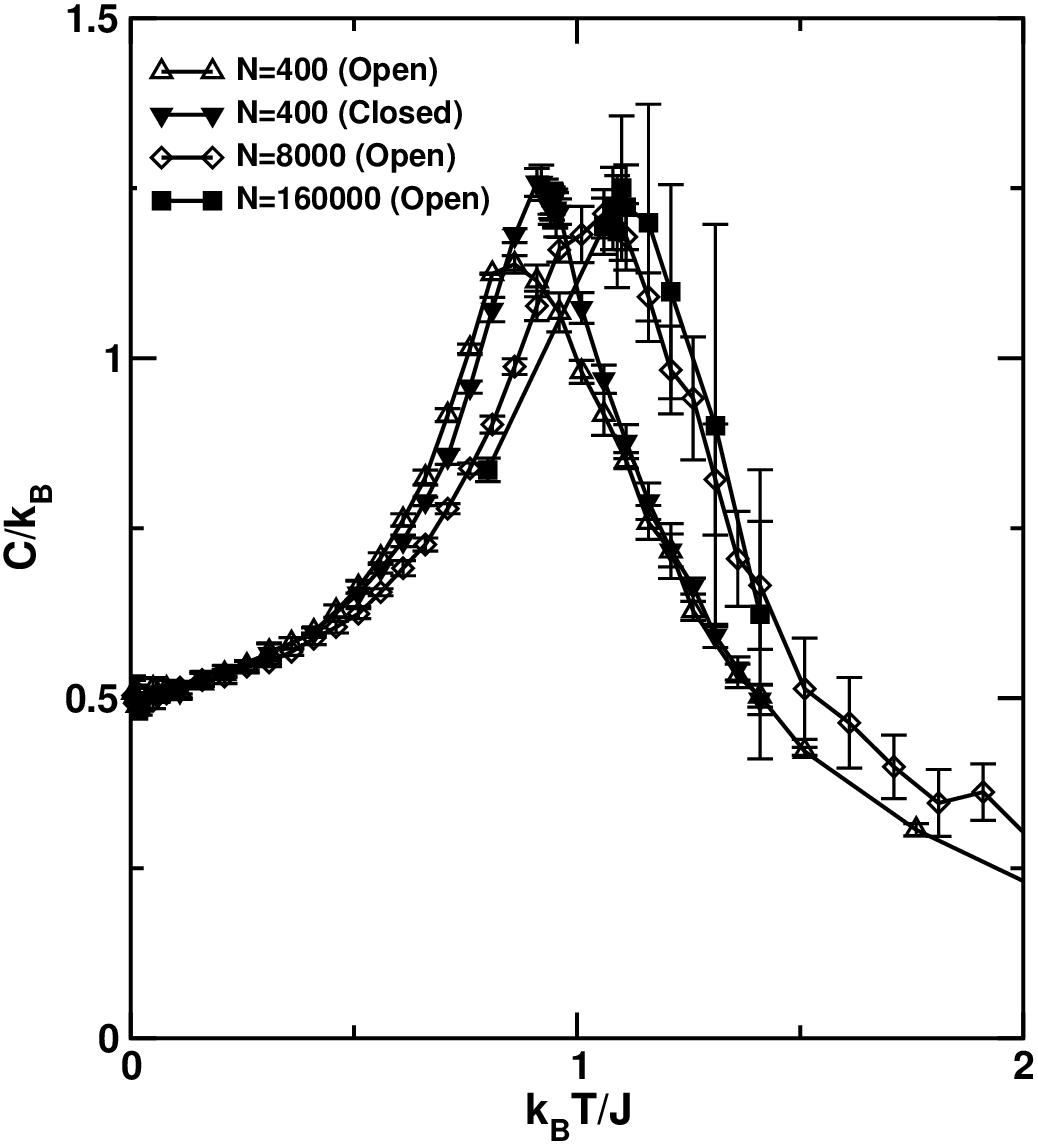}
\end{center}
\caption{Specific heat for Menger sponge with open and closed boundary conditions for different system sizes.
The results for $m=$ 3 Menger sponge ($N=$ 8,000 sites) obtained with closed boundary condition are not shown
for the sake of clarity since they overlap with the results obtained with open boundary condition.}
\label{fig:cv}
\end{figure}

\begin{figure}
\begin{center}
  \includegraphics[angle=0,width=6cm]{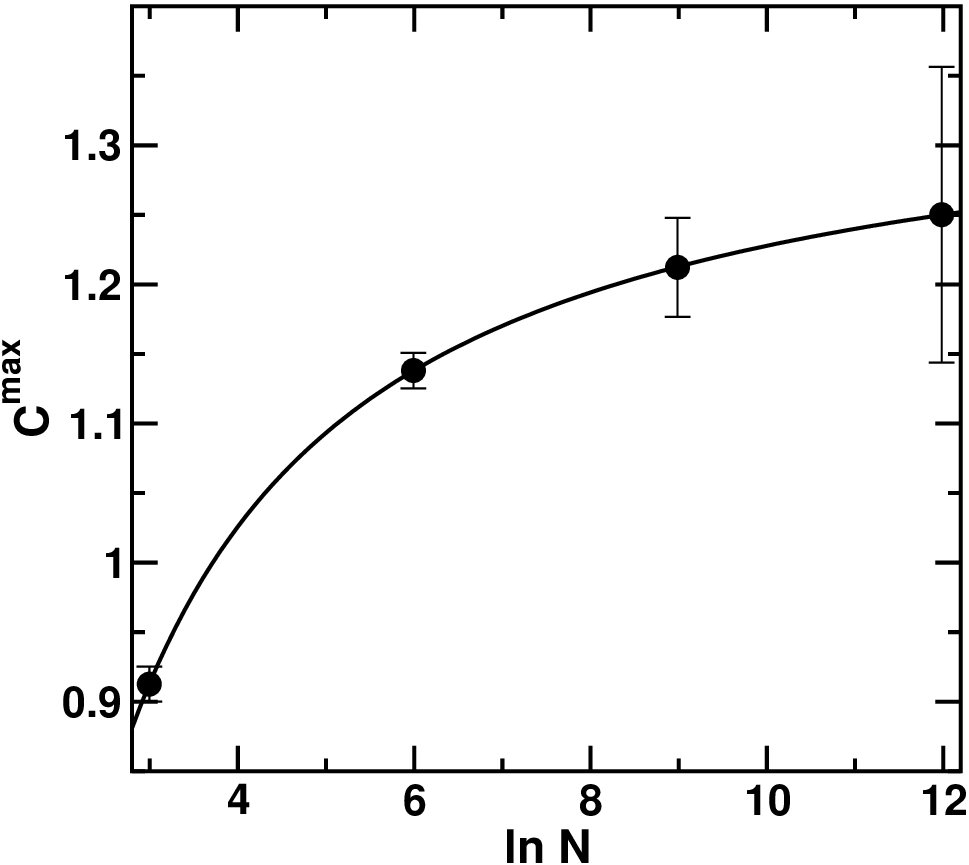}
\end{center}
\caption{The size dependence of the maximum in the specific heat obtained with the open boundary condition
on the system size.}
\label{fig:cvmax}
\end{figure}

Our results for the specific heat are shown in Figure~\ref{fig:cv}. 
In Figure~\ref{fig:cvmax} we show the dependence of the maximum in the specific heat $C^{\mbox{max}}$ obtained 
with the open boundary condition on the system size. The data were fitted by the formula
\begin{equation}
C^{\mbox{max}}=C^{\infty}+\frac{S}{(\ln N)^{a}}\>,
\label{eq:Cvfit}
\end{equation}
and we obtained $C^{\infty}=$ 1.36, $S=$ -1.35, and $a=$ 1.01, with the $\chi^{2}$ of the fit equal 
to 2.03$\times$10$^{-7}$. Thus the maximum in the specific heat of the MS saturates in the 
thermodynamic limit at a value that is not much higher than what we obtained for our largest cluster 
($C^{\mbox{max}}=$ 1.250 $\pm$ 0.106 for $N=$ 160,000). 

The Menger sponge is a structure in three dimensional space but the average coordination number is $\approx$ 4 and is 
in the range characteristic of a two-dimensional system. The specific heat exponent $\alpha$ is not accurately 
known for the classical 
XY-model on periodic lattices in $d=$ 3. The high-temperature series expansions give $\alpha=$ -0.02 $\pm$ 0.03~
\cite{fmw73} and $\alpha=$ 0.02 $\pm$ 0.02~\cite{pjf74}. The experimental results derived from $C_p$ 
measurements~\cite{ahlers73} and from the thermal expansion coefficient measurements~\cite{mpa75} near the 
superfluid transition of $^{4}$He give $\alpha=$ -0.02 $\pm$ 0.02 and $\alpha=$ -0.026 $\pm$ 0.004, respectively. 
Le Guillou and Zinn-Justin~\cite{lz80} have obtained $\alpha=$ -0.007 $\pm$ 0.006 using field-theoretical methods.  
These results seem to favor a small but negative $\alpha$ in which case there is a sharp cusp in the specific 
heat for an infinite system, and in numerical simulations the specific heat would saturate with 
increasing system size. On the other hand, the classical XY-model on periodic lattices   
\noindent in $d=$ 2 leads to 
Berezinskii-Kosterlitz-Thouless (BKT) transition associated with unbinding of topological defects (vortices 
and antivortices), and the specific heat has an unobservable essential singularity at the transition temperature~
\cite{bn79}. The Monte Carlo simulations~\cite{tc79,vc81} give a peak in the specific heat above the BKT transition 
temperature. The peak is  caused by unbinding of vortex clusters~\cite{tc79} and its size saturates 
in the thermodynamic limit~\cite{vc81}. We would like to point out that Kohring {\it et al.}~\cite{ksw86} presented 
Monte Carlo evidence that a continuous phase transition for the classical XY-model on periodic lattices in 
$d=$ 3 is related to unbinding of vortex strings and that the topological defects of this model on the Menger sponge  
are, in principle, vortex strings in addition to vortices in planes parallel to the faces of the sponge. In 
our numerical work we did not keep track of topological defects since we did not have an a priori reason to 
assume that there is a phase transition for the XY-model on the MS.   

However, if there is no phase transition for the XY-model on MS, as we eventually conclude based on the data for 
the helicity modulus, the peak in the specific heat could result from the average energy per site $\langle E \rangle$ 
changing monotonically from a 
value slightly higher than -2J (the average coordination number is slightly less than 4) at low temperatures to 
near zero in disordered paramagnetic phase at high temperatures.

\begin{figure}
\begin{center}
  \includegraphics[angle=0,width=8cm]{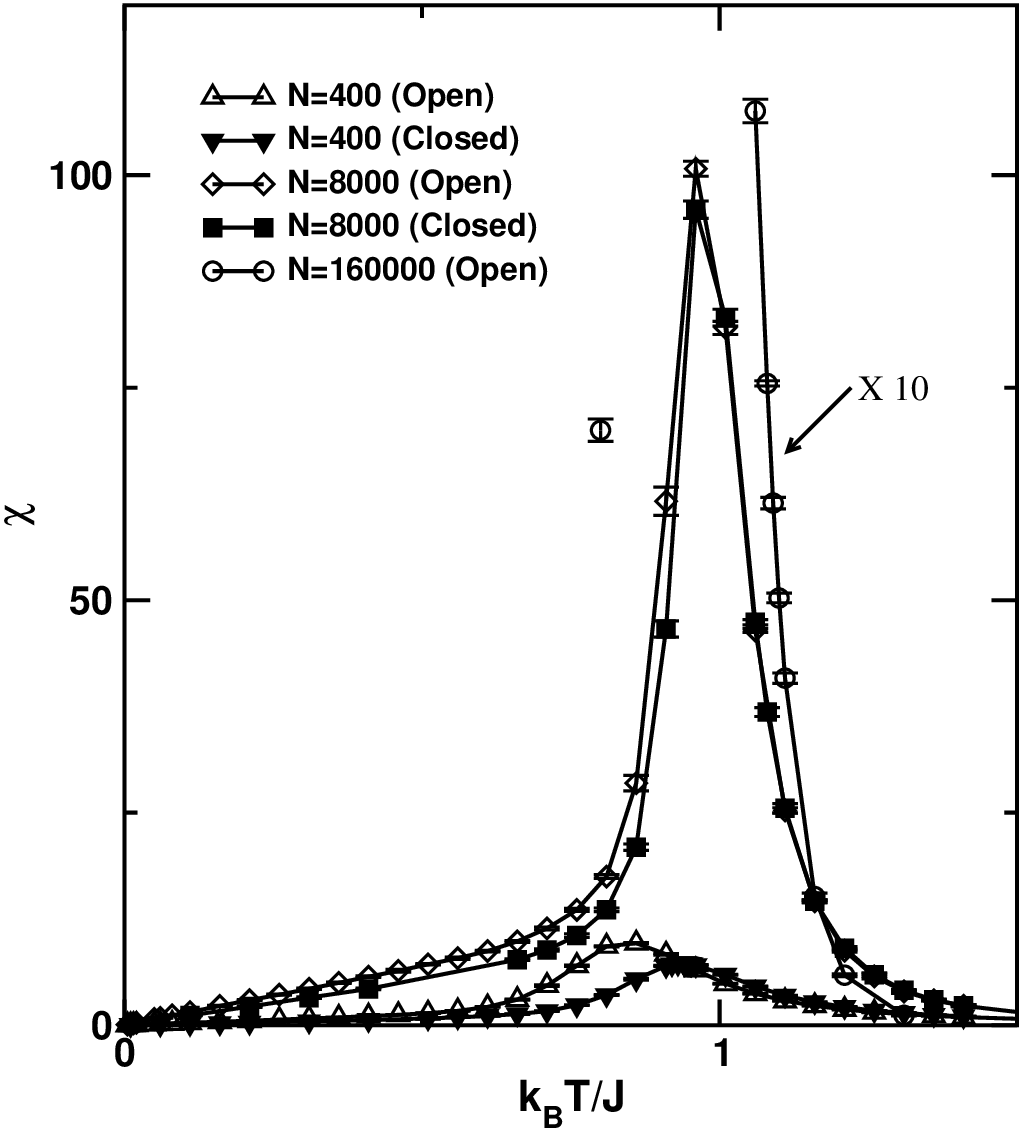}
\end{center}
\caption{Linear susceptibility for Menger sponge with open and closed boundary conditions for different system sizes.}
\label{fig:chi}
\end{figure}

In Figure~\ref{fig:chi} we show our results for the linear susceptibility of the XY-model on the MS 
with different number of sites ($N$) obtained with both types of boundary conditions.
For the classical XY-model on a simple cubic lattice in $d=$ 3 the linear susceptibility $\chi$ diverges at 
transition temperature $k_{B}T_{c}/J$ of about 2.2~\cite{lt89} with the critical exponent $\gamma=$ 
1.316 $\pm$ 0.0025~\cite{lz80}. In numerical simulations on finite cubic lattices with periodic boundary 
conditions one gets a peak in $\chi$ near $k_{B}T_{c}/J=$ 2.2 whose size and sharpness increase with the number of 
sites as a result of a diverging correlation length at the onset of long range order~\cite{pm11}. Also, the 
position of the peak in $\chi$ shows a minute response to the system size, shifting to marginally lower 
temperatures as the size of the lattice is increased~\cite{pm11}. For the MS we find that the size 
of the peak in $\chi$ also increases with the system size, the position of the peak shifts to higher temperatures with 
increasing system size with the open boundary condition, and with the closed boundary condition the position of the 
peak does not change much with the system size for $N\geq$ 400. We note that for the cluster with $N=$ 8,000 sites 
the results for $\chi$ do not depend  
\noindent much on the boundary condition. For the largest cluster with $N=$ 160,000 sites our data at $k_{B}T/J=$ 
0.8 and above $k_{B}T/J=$ 1.06 indicate that $\chi$ has a maximum between these two temperatures, not far from 
$k_{B}T/J=$ 0.95, which is the position of the peak in $\chi$ 
for the sponge with $N=$ 8,000 sites. We note that for the classical XY-model on a square lattice, where the 
theory~\cite{k74} predicts diverging susceptibility $\chi$ above the BKT transition temperature and an infinite  
$\chi$ below the transition temperature, MC simulations give a peak in $\chi$ above the transition temperature whose 
position is much more sensitive to the system size~\cite{mb10} than in the three-dimensional case. In our work on the 
XY-model on fractal lattices Sierpi\' nski gasket~\cite{mb10}, Sierpi\' nski pyramid~\cite{pm11}, and Sierpi\' nski 
carpet~\cite{mp14}, which did not undergo finite temperature phase transition, we found the largest 
sensitivity in $\chi$ to the system size in terms of both peak height and peak position. Therefore our results 
for the susceptibility in Figure~\ref{fig:chi} obtained with the closed boundary condition are not inconsistent 
with a putative continuous phase transition at finite temperature. 

In our previous Monte Carlo work on the classical XY-model on fractal structures~\cite{mb10,pm11,mp14} we found
\begin{figure}
\begin{center}
  \includegraphics[angle=0,width=8cm]{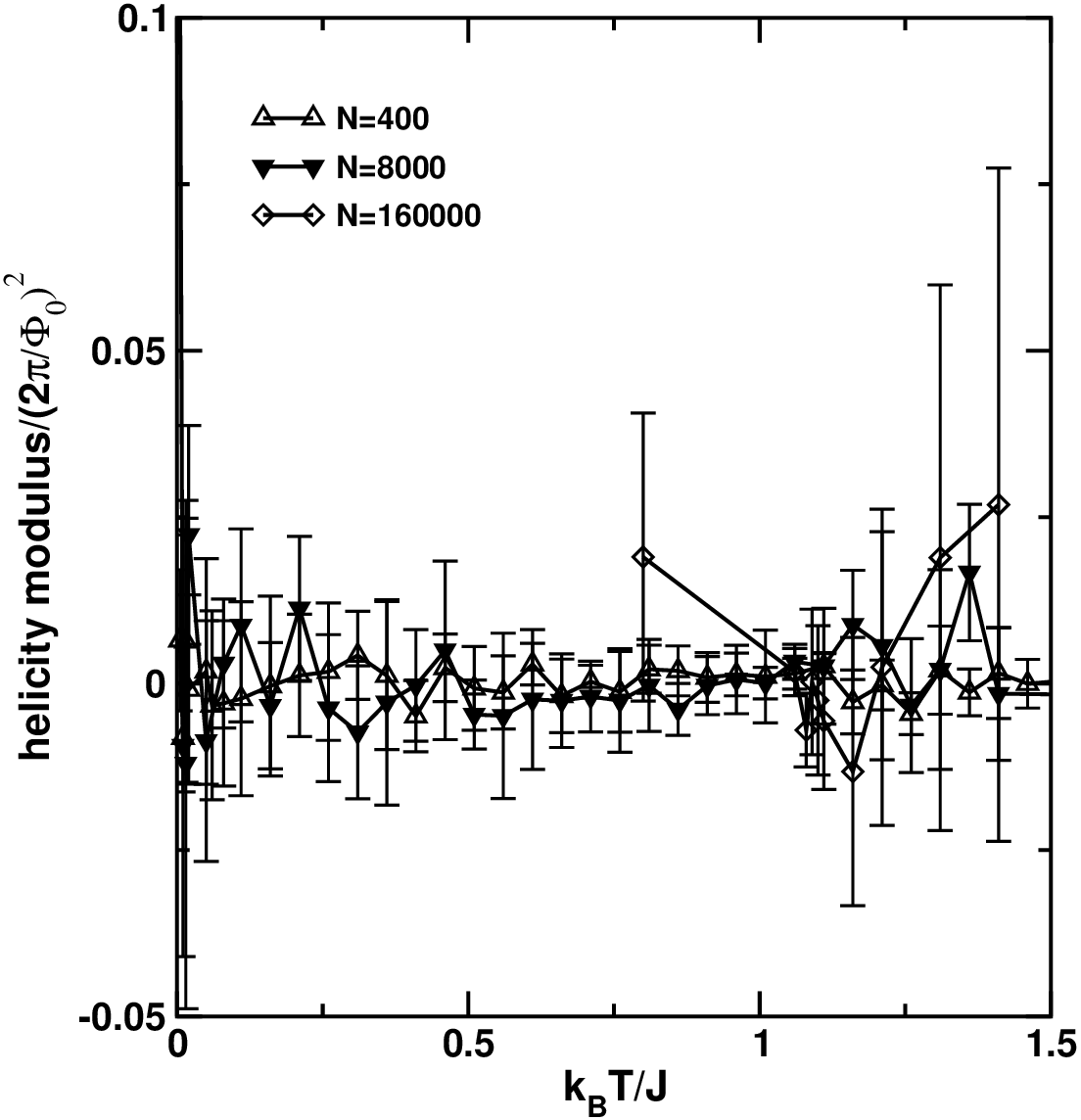}
\end{center}
\caption{Helicity modulus calculated with open boundary condition as a function of the system size. The results for
the first order sponge ($m=$ 1, $N=$ 20) are also zero within the error bars, but are not shown for the sake of
clarity.} 
\label{fig:helOBC}
\end{figure}
the helicity modulus $\gamma(T)$, in particular its dependence on the boundary conditions and on the system size, 
to be a definitive indicator of presence or absence of a finite temperature phase transition. 
$\gamma(T)$ provides a measure of the response of a system to a twisting or helical strain at the 
boundary~\cite{fbj73}. A finite resistance to this twist (finite $\gamma$) is expected for a system at zero
temperature with all the spins aligned. Conversely, in the high temperature paramagnetic phase with 
the spins randomized, this rigidity vanishes.  
In Figure~\ref{fig:helOBC} we show the helicity modulus $\gamma(T)$ calculated with the open boundary condition. 
We find that $\gamma(T)$ vanishes within the error bars for all MS which we considered ($m=$ 1--4). 
The open boundary condition led to vanishing helicity modulus for the XY-model on other fractal structures 
which we considered in our previous work~\cite{mb10,pm11,mp14}. That fact, together with the size dependence 
of the low-temperature $\gamma(T)$ obtained with the closed boundary condition, led us to the conclusion that 
there is no finite temperature phase transition in these systems.
\begin{figure}
\begin{center}
  \includegraphics[angle=0,width=8cm]{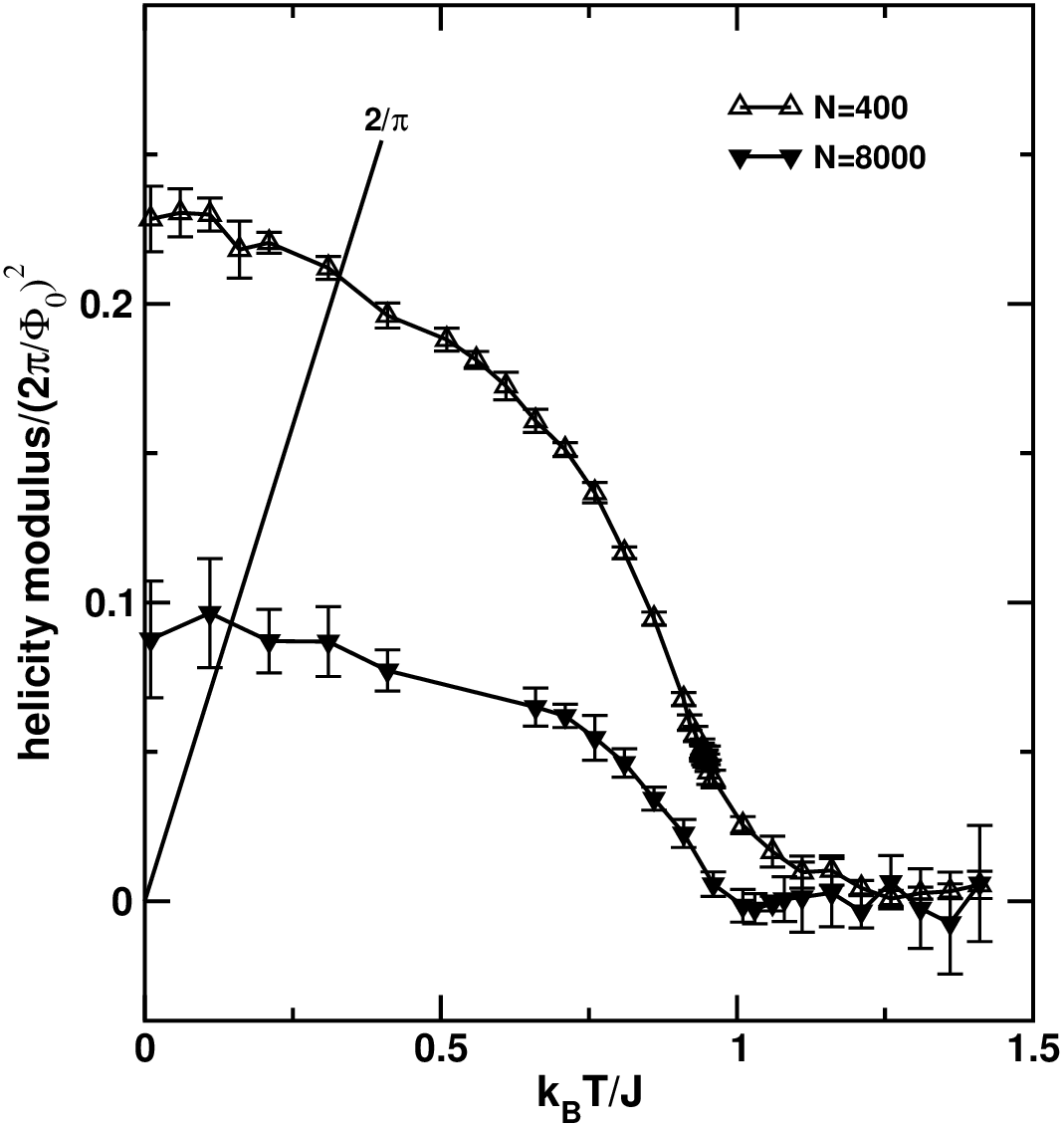}
\end{center}
\caption{Helicity modulus calculated with closed boundary condition as a function of the system size.}
\label{fig:helCBC}
\end{figure}

The results for the closed boundary condition are presented in Figure~\ref{fig:helCBC}. As in the cases of the 
XY-model on Sierpi\' nski gaskets~\cite{mb10}, Sierpi\' nski pyramids~\cite{pm11}, and Sierpi\' nski  
carpets~\cite{mp14}, the closed boundary conditions lead to finite values of the low-temperature   
helicity modulus which decrease with increasing system size. One qualitative difference between the results in 
Figure~\ref{fig:helCBC} and those obtained previously is that for Sierpi\' nski gaskets, Sierpi\' nski pyramids, 
and the Sierpi\' nski carpets the onset of the downturn in $\gamma(T)$ shifted to lower temperatures with increasing 
system size, but it consistently began around the universal $2/\pi$-line for all system sizes (Nelson and 
Kosterlitz~\cite{nk77} predicted a universal jump in $\gamma$ at the BKT transition temperature $T_{c}$ given by
$\gamma(T_{c})/T_{c}=2/\pi$). In the present case, the onset of the downturn in $\gamma$ appears to be 
unrelated to the $2/\pi$-line, occurring at much higher temperatures,   
and the onset does not shift
substantially to lower temperatures with increasing sponge order. The continuous transition that occurs in the  
classical XY-model on regular three-dimensional systems is accompanied by power-law decay of $\gamma$ in the  
vicinity of the transition temperature: $\gamma \propto |T-T_{c}|^{\nu}$ with $\nu = 0.662(7)$~\cite{gh93}
for cubic lattices. Also, the low-temperature values of the helicity modulus do not depend on 
the system size~\cite{lt89}. In MS with the closed boundary condition, the helicity modulus
appears to obey power law decay at temperatures below the temperature where it vanishes.  
\begin{figure}
\begin{center}
  \includegraphics[angle=0,height=8cm]{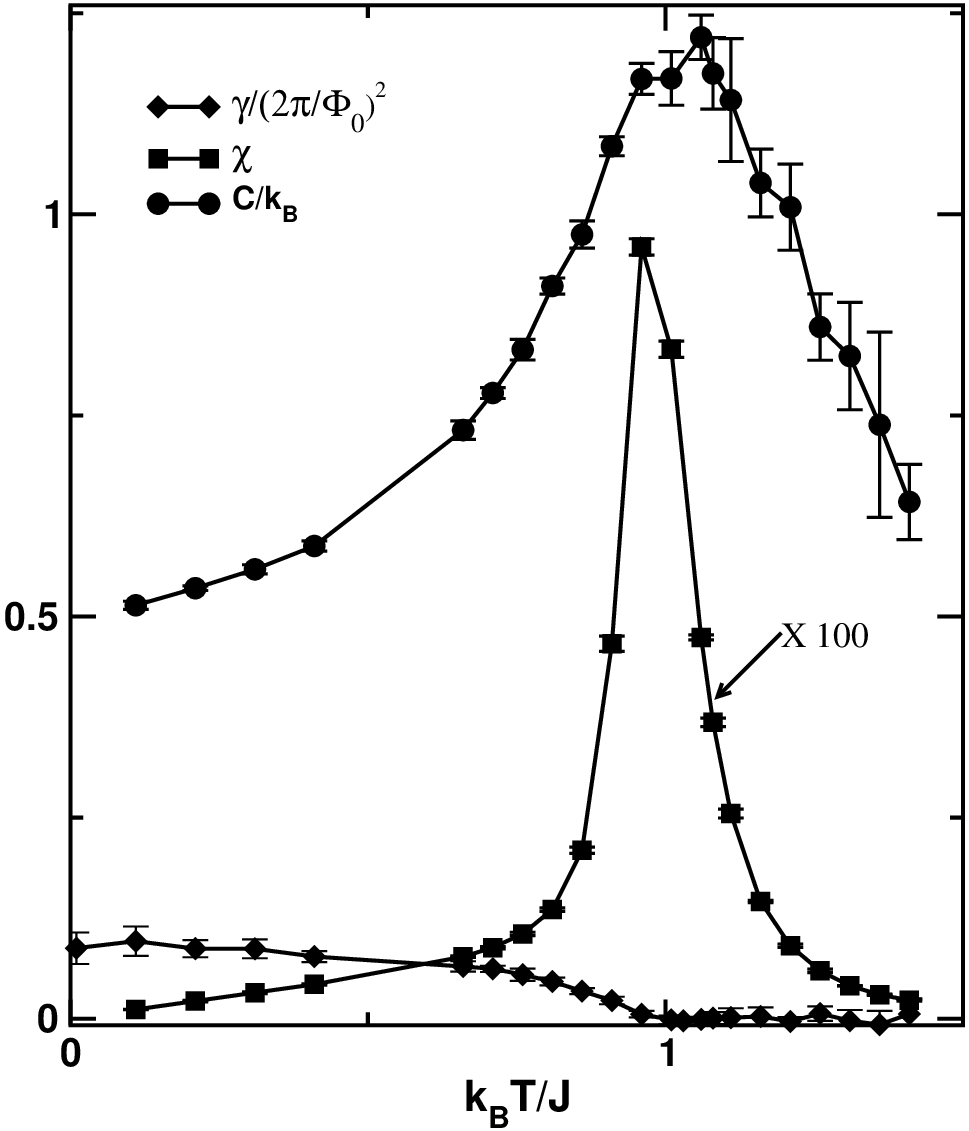}
\end{center}
\caption{Comparison of thermodynamic quantities in the 8000-site Menger sponge with closed boundary conditions.}
\label{fig:comparison}
\end{figure}

Figure~\ref{fig:comparison} shows a comparison of the thermodynamic quantities for the 8000-site sponge 
with closed boundary conditions. One can see that similar to cubic lattice, the heat capacity 
and susceptibility are peaking at roughly the same temperature where the helicity modulus vanishes. 
Taken on their own these results would suggest that there is a putative continuous phase transition at 
finite temperature. 

\begin{figure}
\begin{center}
  \includegraphics[angle=0,width=6cm]{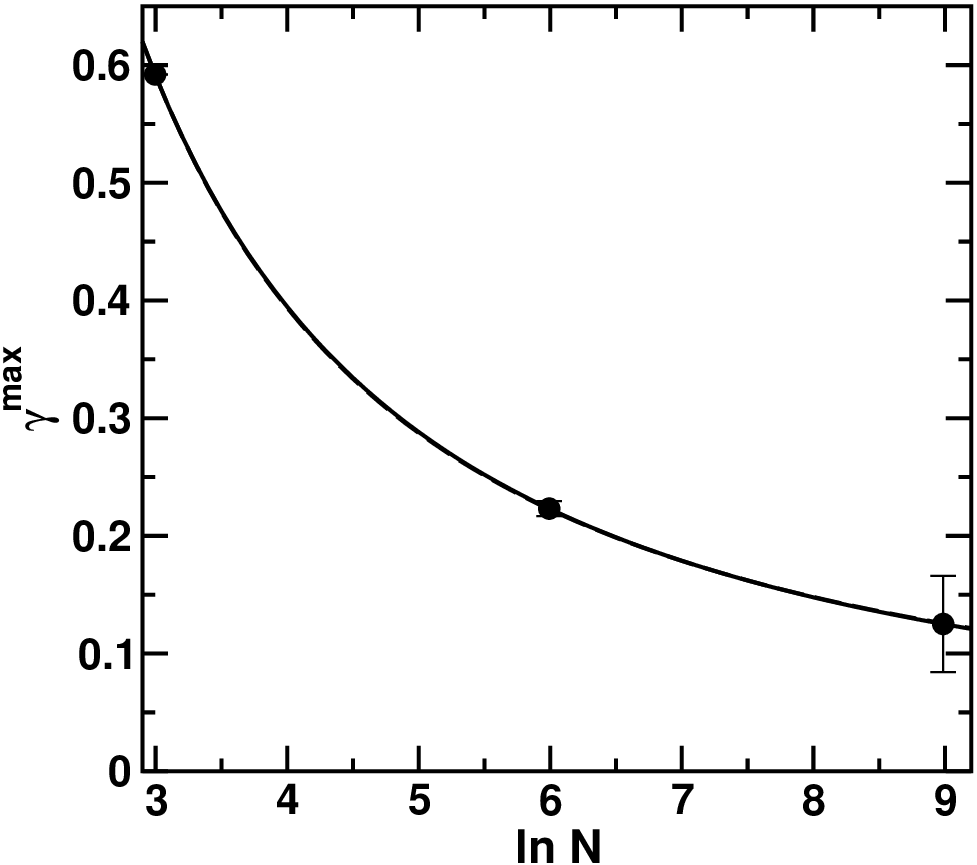}
\end{center}
\caption{The low temperature values of the helicity modulus calculated with closed boundary condition as a 
function of the system size.}
\label{fig:helfit}
\end{figure}

However, the low-temperature values of $\gamma$ for MS depend on the system size. 
In Figure~\ref{fig:helfit} we show the low temperature values of the helicity modulus $\gamma^{\mbox{max}}$ as a 
function of the system size and the fit (solid line) according to the formula
\begin{equation}
\gamma^{\mbox{max}}(N)=X+\frac{Y}{(\ln N)^{f}}\>,
\label{eq:Helfit}
\end{equation}
with $X=$ -0.0045, $Y=$ 2.74, and $f=$ 1.39, and the $\chi^{2}$ of the fit of 2.07$\times$10$^{-7}$. 
These values imply that $\gamma^{\mbox{max}}$ vanishes for $N=$ 6.4$\times$10$^{43}$, which corresponds to 
the thermodynamic limit. A fit 
where $X$ was set equal to 0 was equally good (dashed line in Figure~\ref{fig:helfit} which overlaps with 
the solid line). That fit produced $Y=$ 2.78 and $f=$ 1.41, with $\chi^{2}$ of the fit of 5.72$\times$10$^{-7}$.
These results show that the low-temperature value of the helicity modulus obtained with the closed boundary 
condition goes to zero in the thermodynamic limit $N\rightarrow\infty$. Hence, we have established numerically that 
there is no finite temperature phase transition for the classical XY-model on MS.

We should point out that we reproduced all of these results in simulations on the Menger sponges of orders  
$m=$ 1--3 using the Metropolis MC
algorithm but with slightly larger error bars, except for the heat capacity at high temperatures beyond the maximum 
in $C$, where the Metropolis algorithm produced smaller error bars than the Wolff algorithm. 

\section{Conclusions}\label{summary}

We have performed a Monte Carlo study of the classical XY-model on Menger sponge, which  
has infinite order of ramification, fractal dimension $D=2.7268$, and 
lacunarity $L=0.19204$. The open boundary condition leads to zero helicity modulus at finite temperatures. 
With the closed boundary condition, the low temperature values of $\gamma$ are finite but decrease with increasing 
system size. These trends suggest that the closed boundary conditions introduce additional correlations  
compared to those present in the system with the open boundary condition. By performing the finite-size 
scaling of the low temperature value $\gamma^{\mbox{max}}$ of the helicity modulus obtained with the closed 
boundary condition, we found $\gamma^{\mbox{max}}=$ 0 in the thermodynamic limit $N\rightarrow\infty$.  
Therefore, there is no finite temperature phase transition for the classical XY-model on the Menger sponge, 
despite the fact that it is infinitely ramified, and has fractal dimension
larger than the lower critical dimension of the XY-model. Gefen {\it et al.}~\cite{gab84} pointed out that there is 
no phase transition for spin models with continuous symmetry on a class of Sierpi\' nski carpets with $D<2$. 
They suggested that such models should also be investigated on infinitely ramified fractals of fractal 
dimension $D>2$ (which is the lower critical dimension of the XY-model). Our work represents such a study
and it shows that the value of the fractal dimension D (2.7268 in the present study) cannot be 
the deciding factor in determining whether or not the phase transition takes place.


%
%
%

\end{document}